\documentclass[12pt]{article}
\usepackage{epsf}

\newlength{\dinwidth}
\newlength{\dinmargin}
\def\lapproxeq{\lower .7ex\hbox{$\;\stackrel{\textstyle <}{\sim}\;$}}
\def\gapproxeq{\lower .7ex\hbox{$\;\stackrel{\textstyle >}{\sim}\;$}}

\def\be{\begin{equation}}
\def\ee{\end{equation}}
\def\bea{\begin{eqnarray}}
\def\eea{\end{eqnarray}}

\begin{document}
\titlepage

\begin{flushright}
IPPP/05/12 \\
DCPT/05/24\\
13th April 2005 \\
\end{flushright}

\vspace*{4cm}

\begin{center}
{\Large \bf Diffractive $W+2$~jet production: a background to  \\[2mm]
exclusive $H\to WW$ 
production at hadron colliders}

\vspace*{1cm} \textsc{V.A.~Khoze$^{a,b}$, M.G. Ryskin$^{a,b}$ and W.J. Stirling$^{a,c}$} \\

\vspace*{0.5cm} $^a$ Department of Physics and Institute for
Particle Physics Phenomenology, \\
University of Durham, DH1 3LE, UK \\[0.5ex]
$^b$ Petersburg Nuclear Physics Institute, Gatchina,
St.~Petersburg, 188300, Russia \\[0.5ex]
$^c$ Department of Mathematical Sciences, 
University of Durham, DH1 3LE, UK \\%
\end{center}

\vspace*{1cm}

\begin{abstract}
Central exclusive double diffractive Higgs boson production, $pp\to p \oplus H \oplus p$, is now recognised 
as an important search scenario for the LHC. We consider the case when the Higgs boson decays to two $W$ bosons, one of which may be off-mass-shell, that subsequently decay to the $q\bar q l \nu$ final state. An important background to this is from the QCD process $gg\to Wq\bar q$, where the two gluons are required to be in a $J_z=0$, colour-singlet state. 
We perform an explicit calculation and investigate the salient properties
of this potentially important background process.
\end{abstract}

\newpage

% ===========================================================================
\section{Introduction}

\begin{figure}
\begin{center}
\centerline{\epsfxsize=0.4\textwidth\epsfbox{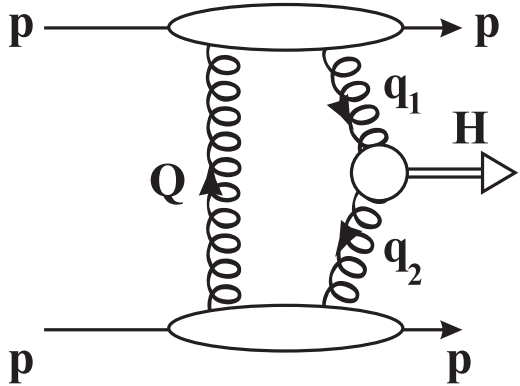}}
\caption{Schematic diagram for central exclusive Higgs production at the LHC,
$pp \to p+H+p$.}  
\label{fig:H}
\end{center}
\end{figure}

Within the last few years the unique environment for investigating new physics
using forward proton tagging at the LHC has become fully appreciated, see for example
\cite{KMRProsp,DKMOR,KKMRext,cox,JE,CR} and references therein.
Of particular interest is the `central exclusive' Higgs boson production $pp\to p \oplus H \oplus p$.
The $\oplus$ signs are used to denote the presence of large rapidity gaps; here
we will simply describe such processes as `exclusive', with
`double-diffractive' production being implied. In these exclusive processes there is no
hadronic activity between the outgoing protons and the decay products of the central system.
The predictions for exclusive production are obtained by calculating the diagram of Fig.~\ref{fig:H}
using perturbative QCD \cite{KMR,KMRProsp}. In addition, we have to calculate and include the probability
that the rapidity gaps are not populated by secondary hadrons from the underlying event \cite{KMRsoft} .

There are three reasons why central exclusive production is so attractive.
First, if the outgoing protons remain intact and scatter through small angles then, to a very good approximation,
the primary active di-gluon system obeys a $J_z=0$, CP-even selection rule
\cite {Liverpool,KMRmm}. Here $J_z$ is the projection of the total angular momentum
along the proton beam axis. This selection rule readily permits a clean determination 
of the quantum numbers of the observed Higgs-like resonance which will be dominantly produced in a scalar state.
Secondly, because the process is exclusive, the energy loss of the outgoing protons is directly
related to the mass of the central system, allowing a potentially excellent mass resolution, irrespective 
of the decay mode of the produced particle.\footnote{Recent studies suggest that the missing mass resolution 
$\sigma$ will be of order $1\%$ for a $140$~GeV 
central system, assuming both protons are detected at 420m from the interaction point \cite{RO,cox}.}
And, thirdly, a signal-to-background ratio of order 1
(or even better) is achievable, even with a moderate luminosity of 30~fb$^{-1}$ \cite{DKMOR,cox}.
In some MSSM Higgs scenarios central exclusive production provides
an opportunity for lineshape analysing \cite{KKMRext,JE} and offers a way
for direct observation of a CP-violating signal in the Higgs sector \cite{KMRCP,JE}.
The analysis in \cite{KMR,DKMOR,KKMRext} was focused primarily
 on light SM and MSSM Higgs production, with the Higgs 
decaying to 2 $b-$jets. The potentially copious $b-$jet (QCD) background is controlled by
 a combination of the spin-parity 
selection rules \cite{Liverpool,KMRmm}, which 
strongly suppress  leading-order  $b \bar b$ production, and the 
mass resolution from the forward proton 
detectors. The missing mass resolution is especially critical in controlling the background,
since poor resolution would allow more background events into the mass window around the resonance.
Assuming a mass window $\Delta M \sim 3 \sigma \sim 3-4$~GeV, it is estimated that 
11 signal events, with a signal-to-background ratio of order 1, can be achieved 
with a luminosity of 30 fb$^{-1}$ in the $b \bar b$ decay
channel \cite {KMRmm,DKMOR}.\footnote{It is worth noting that certain 
regions of MSSM parameter space can be especially `proton tagging friendly'.
For example, at large  $\tan\beta$   
the situation becomes exceptionally 
favourable, with predicted Higgs signal-to-background ratios in excess 
of 20 \cite{KKMRext}. In this particular case the tagged proton mode may well
be {\it the} discovery channel.} 
Whilst the $b \bar b$ channel is theoretically very attractive,
allowing direct access to the dominant decay mode
of the light Higgs boson, there are some basic 
problems which render it challenging from an experimental perspective,
see \cite{ww} for more details.
First, it relies heavily on the quality of
 the mass resolution from the proton taggers to suppress 
the background. 
Secondly, triggering on the relatively low-mass dijet signature of the
$H \rightarrow b \bar b$ events is a challenge
for the Level 1 triggers of both ATLAS and CMS. And, thirdly, this
measurement requires double $b-$tagging, with a corresponding price to pay for
tagging efficiencies.
In Ref.~\cite{ww}, attention was turned to
 the $WW^*$ decay mode of the light Higgs Boson, and above 
 the 2 $W$ threshold, the $WW$ decay mode.\footnote{Note that the rate of detectable events from $H\to ZZ$ decay is
very low --- less than $10\%$ of the $H\to WW$ rate --- and we shall not
consider this channel further here.}  This channel 
does not suffer from any of the above problems: suppression 
of the dominant backgrounds does not rely so strongly on the mass resolution of the 
detectors, and, certainly, in the semi-leptonic decay channel of the $WW$ system 
Level 1 triggering is not a problem. The advantages of forward proton tagging are, however, 
still explicit. Even for the double leptonic decay channel (i.e. with two leptons and two
final state neutrinos), the mass resolution will be very good, and of course 
observation of the Higgs in the double tagged channel immediately 
establishes its quantum numbers. It is worth mentioning
that the mass resolution should improve with increasing Higgs mass \cite{RO}.  
Moreover, the semileptonic `trigger cocktail' may allow
the combination of signals not only from $H\to WW$ decays but also from
the $\tau\tau$, $ZZ$ and even the semileptonic $b-$decay channels.

\begin{figure}
\begin{center}
\centerline{\epsfxsize=0.8\textwidth\epsfbox{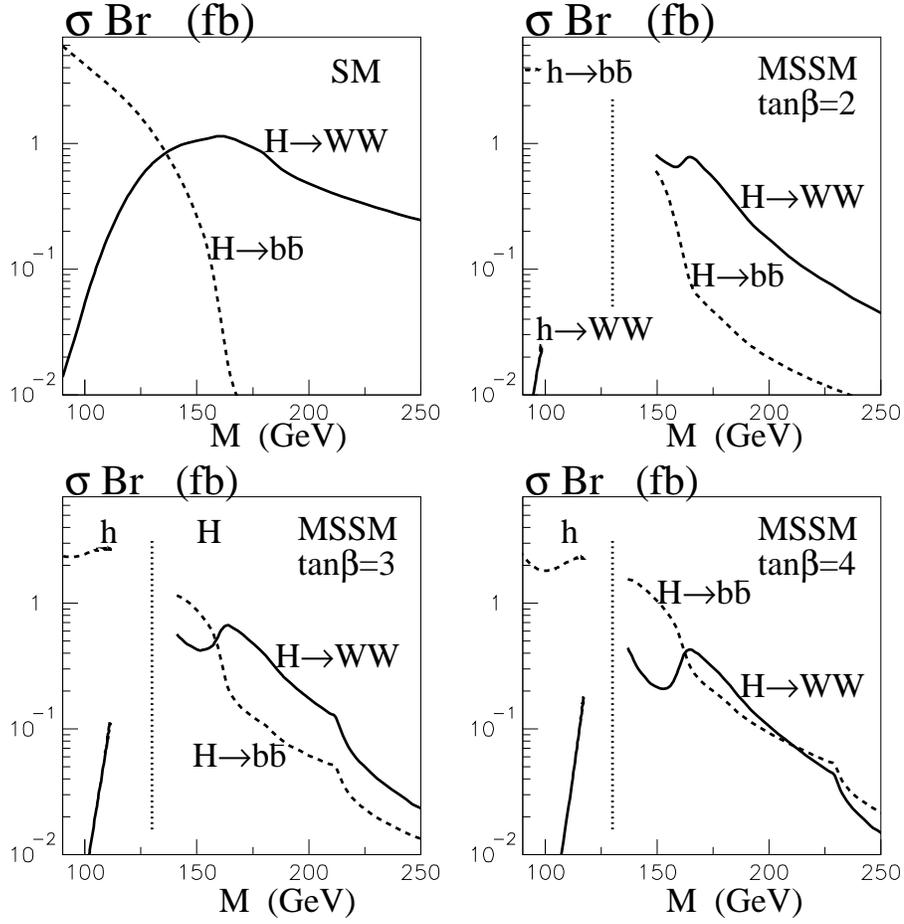}}
\caption{The cross section times branching ratio 
for the central exclusive production of the MSSM Higgs boson (with three values of
$\tan\beta = 2,3,4$)
as a function of Higgs mass in the $WW$ and $b \bar b$ decay channels. The cross section
for Standard Model Higgs boson production is also shown.}  
\label{fig:tanbeta}
\end{center}
\end{figure}

The central exclusive production cross section 
for the Standard Model Higgs boson was calculated in \cite{KMR,KMRProsp}. 
In Fig.~\ref{fig:tanbeta} we show the cross section for 
the process $pp \rightarrow p H p \rightarrow p WW p$ 
as a function of the Higgs mass $M_H$ at the LHC. The increasing 
branching ratio 
to $WW^{(*)}$ (from $12 \%$ at $M_H = 120$~GeV to $\sim 100 \%$ at $160$~GeV)
 as $M_H$ increases (see for example \cite{CH}) 
compensates for the falling central exclusive production cross section. 
For comparison, we also show the cross 
section times branching ratio for $pp \rightarrow p H p \rightarrow p b \bar b p$.
Here, and in what follows, we use version 3.0 of the HDECAY code \cite{HDEC}.
For reference purposes, the cross sections in Fig.~\ref{fig:tanbeta}  are normalised in such a way
that $\sigma_H = 3$~fb for $M_H = 120$~GeV. 

Note also that nowadays there is  renewed interest in MSSM scenarios
with low $\tan\beta$. This is because the most recent value of the top
quark mass weakens the low $\tan\beta$ exclusion bounds
from LEP (see for example \cite{tanbeta}),
and the experimental coverage of this range of the MSSM
parameter space becomes more attractive.
In  Fig.~\ref{fig:tanbeta} we show the results for $\tan\beta=2,3,4$. Evidently the
expected central exclusive diffractive production yield is promising in the low  $\tan\beta$ region.

Experimentally, events with two $W$ bosons in the final
 state fall into 3 broad categories  --- fully-hadronic,
semi-leptonic and fully-leptonic --- depending on 
the decay modes of the $W$'s. Events in which at 
least one of the $W$s decays in either the electron or muon channel 
are by far the simplest, and Ref.~\cite{ww} focuses mainly
on these semi- and fully-leptonic modes. 
As mentioned above, one of the attractive features of the $WW$ channel 
is the absence of a relatively large 
irreducible  background, {\it cf.} the large 
central exclusive $b \bar b$ QCD background  in the case of $H \rightarrow b \bar b$,
suppression of which
relies strongly on the experimental missing mass resolution 
and di-jet identification.

The primary  exclusive  backgrounds in the case of the $WW$ channel
 can be divided into two broad categories: 
\begin{enumerate}
\item Central production of a $WW^*$ pair $pp\to
p+(WW^*)+p$ from either the (a) $\gamma\gamma\to WW^*$ or (b)
$gg^{PP}\to WW^*$ subprocess.
\item 
The $W$-strahlung process  $pp\to p+Wjj+p$
originating in the $gg^{PP}\to Wq \bar q$
subprocess, where  the $W^*$ is `faked' by the
two quarks.
\end{enumerate}
Here the notation $gg^{PP}$ indicates that each active gluon
comes from a colour-singlet $t-$channel (Pomeron) exchange
and that the di-gluon state obeys the $J_z=0$, parity-even selection
rule. 
As shown in \cite{ww}, over a wide region of  Higgs masses
the photon-photon backgrounds are strongly suppressed
if we require that the final leptons and jets are central
and impose cuts on the transverse momenta of the protons in the taggers.
Using the results of Ref.~\cite{PM}, we find that the QCD quark-box-diagram contribution
from the $gg^{PP}\to WW^*$ subprocess
is also very small, on the level of  $1\%$ of the signal yield.
The most important  background therefore comes from the second category above, i.e. from
the $W$-strahlung process exemplified by the diagrams of Fig.~\ref{fig:Wqq}.
Here
we have to take into account the $J_z=0$ projection of this amplitude, which requires
a calculation of the individual helicity amplitudes.  This can be done, for example, 
using the spinor technique of Ref.~\cite{kwjs}.
The primary aim of this paper is to investigate the salient properties
of this potentially important background process.

\section{The $gg\to Wq\bar q$ $J_z=0$ colour-singlet hard process}

The tree-level ${\cal O}(\alpha_S^2\alpha_W)$ process $gg\to Wq\bar q$ is one of many processes that contribute
to $W+2$~jet production at hadron colliders, and as such it has been studied intensively as part of the `QCD' 
background to $W^+W^-$, $t \bar t$ etc. production. The scattering amplitude was first calculated almost 
twenty years ago in Ref.~\cite{kwjs}, using the new (at that time) spinor techniques that were developed for multiparton tree-level scattering amplitudes. Nowadays, all the spin- and colour-summed amplitudes contributing
to inclusive $W+2$~jet production are easily obtained from automated tree-level matrix element software packages such as MADGRAPH \cite{MADGRAPH}.

However in the present context we are specifically interested in the $J_z=0$, colour-singlet projection of
the $gg\to Wq\bar q$ process. It is difficult to extract such projections from the standard packages, and so
we have performed the calculation from first principles using the original techniques of Ref.~\cite{kwjs}.

\begin{figure}
\begin{center}
\centerline{\epsfxsize=0.4\textwidth\epsfbox{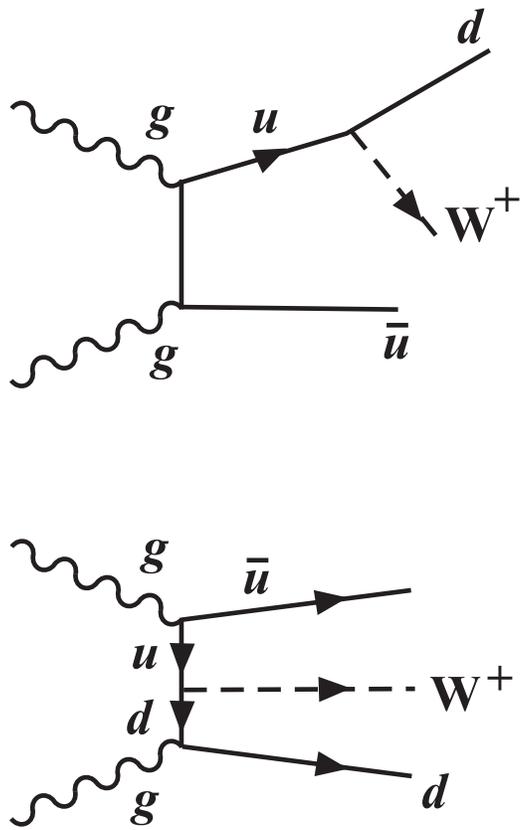}}
\caption{Examples of Feynman diagrams contributing to the $J_z=0$, colour-singlet $gg\to Wq\bar q$ process.}  
\label{fig:Wqq}
\end{center}
\end{figure}

It is interesting to compare the structure of the inclusive and projected amplitudes. For the former, there are
a total of 8 Feynman diagrams, two involving the triple-gluon vertex ($gg\to g^* \to Wq\bar q$) and six diagrams
corresponding to the six different permutations of the three gauge bosons attached to the quark line. 
Two of the latter are shown in Fig.~\ref{fig:Wqq}. Three of the six diagrams correspond to an interchange of the two
gluons, and so the sum and difference of these amplitude triplets (labelled $A_{123}$ and $A_{456}$) contribute to even and odd colour factors respectively.
Schematically, then, we have for the inclusive case:
\begin{eqnarray}
\vert {\cal M}\vert^2(J_z=0) & = &  \frac{28}{3}   \left(  \vert A_{123}(++) + A_{456}(++)  \vert^2
 +  \vert  A_{123}(--) + A_{456}(--)\vert^2      \right)     \nonumber \\
& + & 12 \left(  \vert A_{123}(++) - A_{456}(++) +2A_{78} \vert^2
 +  \vert  A_{123}(--) - A_{456}(--)  +2A_{78} \vert^2   \right) \nonumber \\
\vert {\cal M}\vert^2(J_z=2) & = &  \frac{28}{3}   \left(  \vert A_{123}(+-) + A_{456}(+-)  \vert^2
 +  \vert  A_{123}(-+) + A_{456}(-+)\vert^2      \right)     \nonumber \\
& + & 12 \left(  \vert A_{123}(+-) - A_{456}(+-) \vert^2
 +  \vert  A_{123}(-+) - A_{456}(-+)\vert^2   \right) 
\label{eq:amp1}
\end{eqnarray}
where the labels $(++)$ etc. are the helicities of the incoming gluons (the quark helicities being fixed once
the sign of the $W$ boson is specified), and the even and odd colour factors are
\begin{eqnarray}
\frac{28}{3}& = & {\rm Tr}[(T^aT^b+T^bT^a)(T^aT^b+T^bT^a)] \nonumber \\
12 & = & {\rm Tr}[(T^aT^b-T^bT^a)(T^bT^a-T^aT^b)]
\end{eqnarray}
Note that the diagram with the $s-$channel gluon contributes only to the colour-odd, $J_z=0$ amplitude. In the fully
inclusive $W+2$~jet calculation, the two spin contributions of Eq.~(\ref{eq:amp1}) are of course added together.

For the background to exclusive $H\to WW$ production, Fig.~\ref{fig:H}, we need the $J_z=0$, colour-singlet projection:
\begin{equation}
\vert {\cal M}\vert^2(J_z=0, \mbox{colour singlet}) =  \frac{64}{3}   \left(  \vert A_{123}(++) + A_{456}(++) 
+  A_{123}(--) + A_{456}(--)\vert^2      \right)
\label{eq:amp2}
\end{equation} 
where now the colour factor is 
\begin{equation}
\frac{64}{3} =  4\;{\rm Tr}[T^aT^aT^bT^b]
\end{equation}
The colour-octet $s-$channel gluon diagrams ($A_{78}$) no longer contribute to the $J_z=0$ amplitude.
Note also that the $(++)$ and $(--)$ spin components can interfere in the overall amplitude squared. This is because
in the case of exclusive central diffractive production, the amplitudes (rather than the cross sections) should be averaged over the helicities $(++)$ or $(--)$  of the incoming gluons, see for example \cite{KMRProsp}.

In the following section we present some numerical results based on the above calculations.

\section{Numerical results and discussion}

\begin{figure}
\begin{center}
\centerline{\epsfxsize=0.7\textwidth\epsfbox{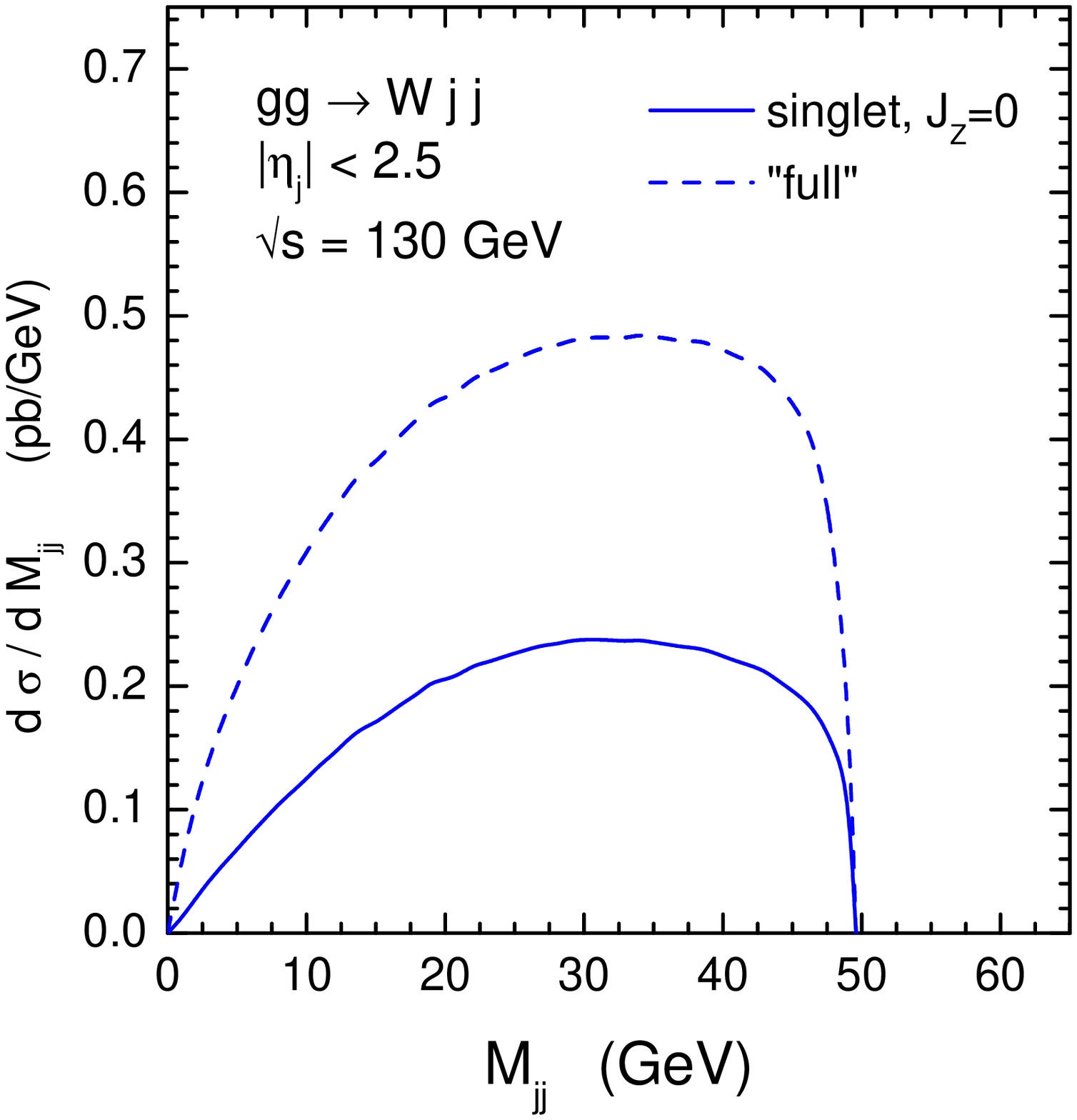}}
\vspace{-5cm}
\caption{The $J_z=0$, colour-singlet $gg\to Wq\bar q$ cross section of Eq.~(\ref{eq:amp2}) compared to the spin- and colour-summed cross section of Eq.~(\ref{eq:amp1}). The $gg$ centre-of-mass energy is $\sqrt{s} = 130$~GeV, and the 
final-state quark jets  are required to lie in the pseudorapidity range $-2.5 < \eta_j < +2.5$.}  
\label{fig:ggWW}
\end{center}
\end{figure}

\begin{figure}
\begin{center}
\centerline{\epsfxsize=0.7\textwidth\epsfbox{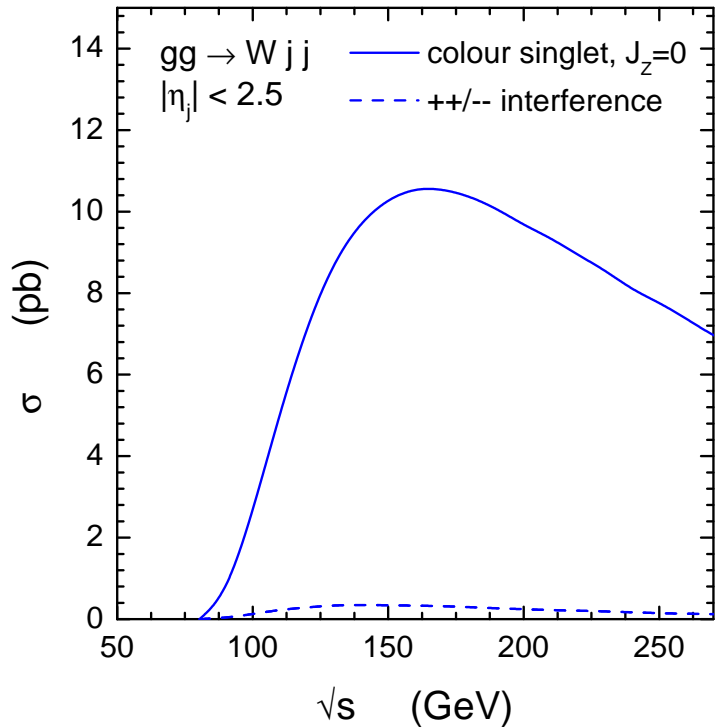}}
\vspace{-5cm}
\caption{The scattering energy dependence of the $J_z=0$, colour-singlet $gg\to Wq\bar q$ cross section of 
Eq.~(\ref{eq:amp2}). The final-state quark jets  are required to lie in the pseudorapidity range 
$-2.5 < \eta_j < +2.5$. Also shown (dashed line) is the contribution of the interference between 
the $(++)$ and $(--)$ gluon helicity amplitudes in Eq.~(\ref{eq:amp2}).}  
\label{fig:ggWWtot}
\end{center}
\end{figure}

\begin{figure}
\begin{center}
\centerline{\epsfxsize=0.7\textwidth\epsfbox{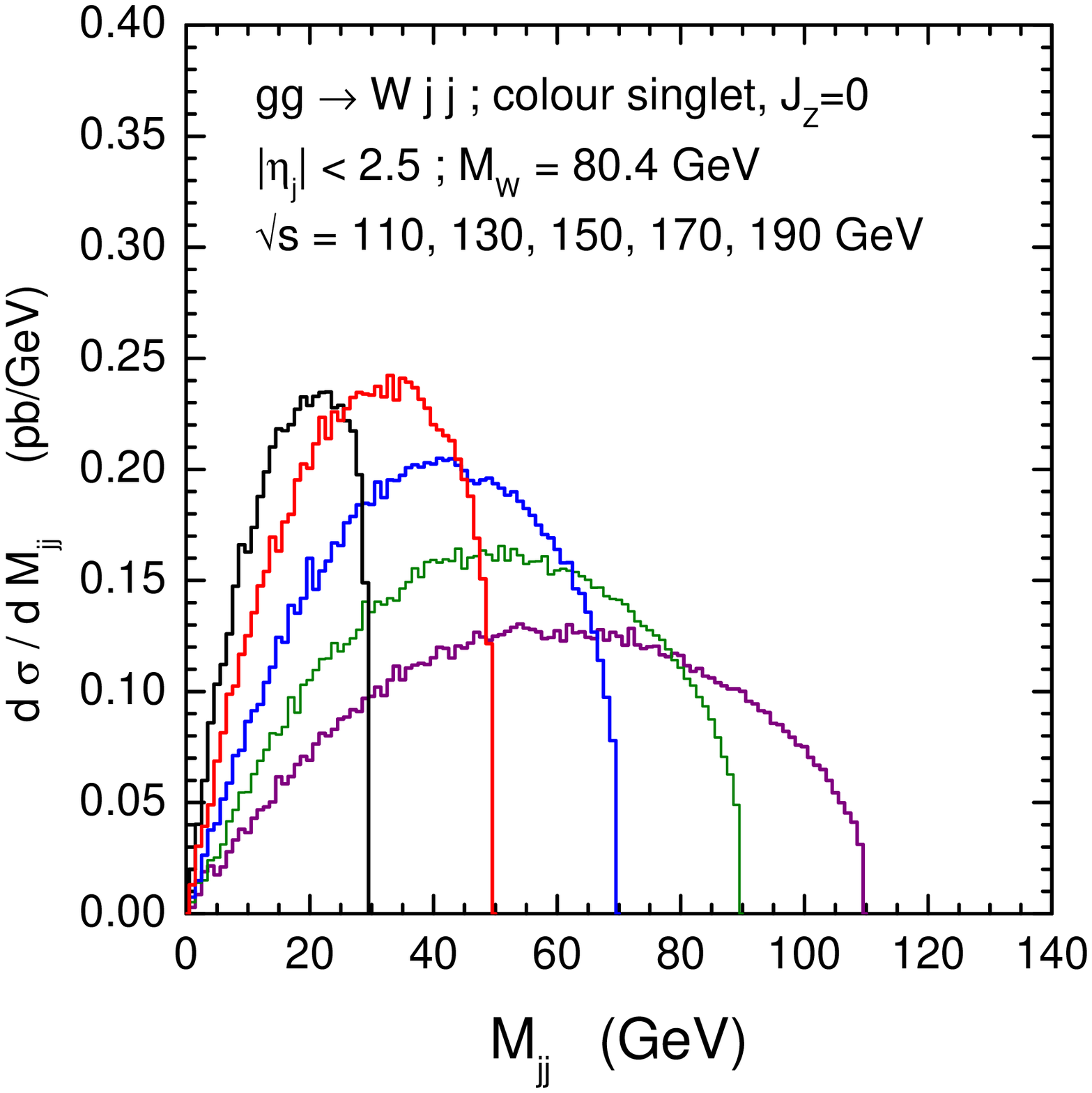}}
\vspace{-5cm}
\caption{The jet-jet ($q\bar q$) invariant mass distribution for  $J_z=0$, colour-singlet $gg\to Wq\bar q$ production for 
different values of the $gg$ centre-of-mass energy $\sqrt{s}$. The final-state quark jets  are required to lie in the pseudorapidity range $-2.5 < \eta_j < +2.5$. The end point for each distribution is at $M_{jj} \approx \sqrt{s}-M_W$. }  
\label{fig:ggWWmjj}
\end{center}
\end{figure}

The $gg^{PP}\to Wq\bar q$ cross section is obtained by integrating the matrix element squared
of the previous section over an appropriate region of three-body phase space. However care must be taken
in avoiding collinear singularities. These occur when either one or both of the incoming gluons splits into a collinear $q\bar q$ pair. Of course in this case the final-state (zero transverse momentum) 
quarks would not be registered as jets in the detector. In order to have observable jets and to suppress the collinear
logarithmic singularities, we impose
the pseudorapidity cut ($|\eta_{j}|<2.5$) on the final-state quarks. With these minimal cuts, we obtain a finite background cross section with which to compare the Higgs signal.

Figure~\ref{fig:ggWW} shows the jet-jet mass distribution $d \sigma/d M_{jj}$ for the inclusive and $J_z=0$, colour-singlet projected $gg\to Wq\bar q$ process at $\sqrt{s} = 130$~GeV, a typical value for the Higgs mass. 
Two families of fermions are summed over in the final state (i.e. the quarks are either $u$, $d$, $s$ or $c$) and both 
 $W^+q\bar q$ and $W^-q\bar q$ configurations are included. Evidently the  $J_z=0$, colour-singlet projection suppresses the cross section by about a factor of 2 for these kinematics. In Fig.~\ref{fig:ggWWtot} we show the total  $gg^{PP}\to Wq\bar q$ cross section as a function of the gluon-gluon centre-of-mass energy $\sqrt{s}$. The dashed line is the contribution of the interference between the $(++)$ and $(--)$ gluon helicity amplitudes in Eq.~(\ref{eq:amp2}). This is evidently a very small effect for these kinematics. Finally, Fig.~\ref{fig:ggWWmjj} shows the jet-jet ($q\bar q$) invariant mass distribution for different values of $\sqrt{s}$.

{}From Fig.~\ref{fig:ggWWtot} we see that the $gg\to Wq\bar q$ total cross section  is about $7.2\; (9.8)$~pb for $\sqrt{s}=120\; (140)$~GeV, rising to  
 10.6~pb  at $\sqrt{s}\simeq 160$~GeV and then decreasing slowly for higher energies. When comparing to the Higgs $\to W W^{(*)}$  or $WW$ signal, this background cross section
should be multiplied by the phase space factor  $2\Delta M/M_H$,
where $\Delta M\sim 3\sigma$ is the mass window over which we collect the signal,
and by the corresponding gluon luminosity at $\sqrt{s_{gg}} = M_H$ \cite{KMRProsp}.
Assuming that 
 $2\Delta M/M_H\sim 0.1$
we finally arrive at the 
background cross section at $\sqrt{s}=14$~TeV of about 1.7~fb for
$M_H=140$~GeV. This is a `maximal' background cross section in the sense that the only cuts on the final-state jets --- apart from the mass-window requirement --- are the weak rapidity cuts $\vert \eta_j\vert < 2.5$ imposed in the jet-jet centre of mass system. Further laboratory-frame cuts on jet and lepton rapidity and transverse momentum will further reduce the background cross section, but will of course also reduce the signal, though to a lesser extent. Comparing with Fig.~\ref{fig:tanbeta}, we see that the $Wq\bar q$ background cross section  is about a factor of two larger
than the Standard Model Higgs signal\footnote{Strictly, in this comparison the signal cross sections in 
Fig.~\ref{fig:tanbeta}
should be reduced slightly to take into account the $W\to \sum q\bar q$ branching ratio and the rapidity acceptance of the jets.} at this value of $M_H$. 

\begin{figure}
\begin{center}
\centerline{\epsfxsize=0.7\textwidth\epsfbox{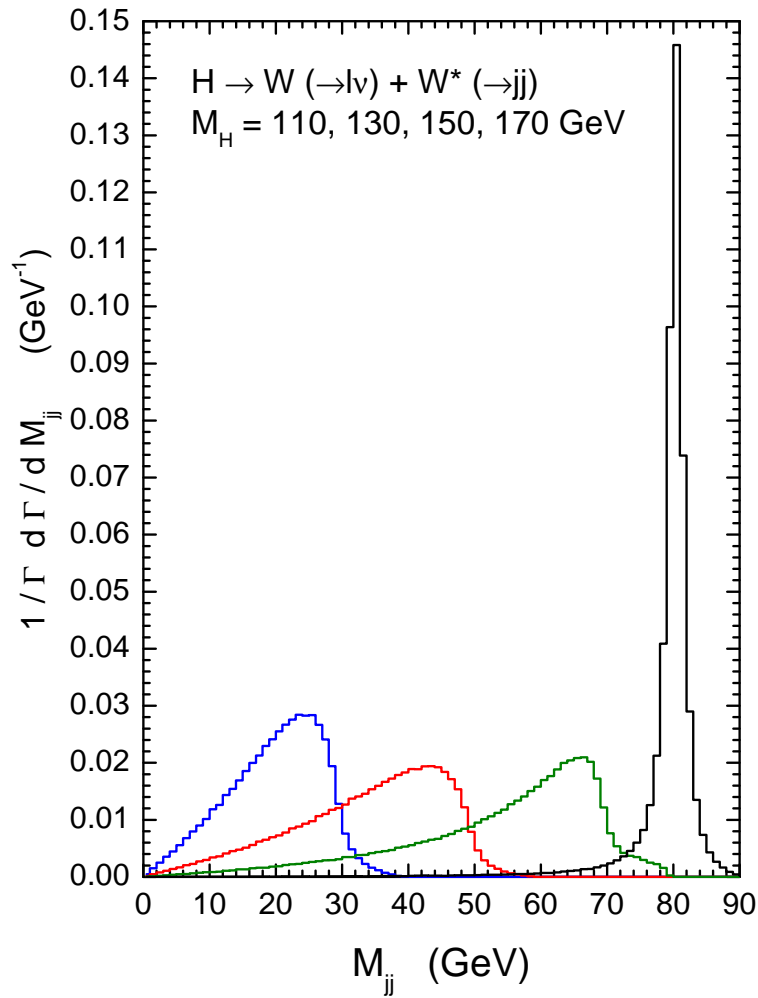}}
\vspace{-2cm}
\caption{The jet-jet ($q\bar q$) invariant mass distribution for the signal process $H \to W(\to l \nu) + W^*(\to q \bar q)$, for different values of the Higgs boson mass. For $M_H > 2 M_W$, the jet-jet mass distribution assumes a Breit-Wigner form. The distribution integrates to 0.5 in each case.}  
\label{fig:Hmjj}
\end{center}
\end{figure}

It might appear that the QCD background could be further reduced by only selecting the subset of
events with a rather large two-jet mass $M_{qq}$, in order to mimic the $ W^{(*)} \to q \bar q$ signal. However as shown in Fig.~\ref{fig:Hmjj}, when $M_H < 2 M_W$ and one of the $W$ bosons is off-mass-shell, the $M_{W^*}$ distribution is peaked {\it below} the edge of phase space at $M_{W^*}\simeq M_H-M_W$. The reason for this  (see, for example,
Ref.~\cite{SK}) is that below the nominal $WW$ threshold
the three-body phase space factor compensates the variation
of the $W$ Breit-Wigner distribution in the tail. Indeed, the $M_{W^*}$ distribution actually vanishes
 at $M_{W^*}=M_H-M_W$ (where here $M_H$ denotes the c.m.s. energy of the
$WW^*$ system). On the other hand, for a low mass $M_{W^*}<M_W$ the
variation of the Breit-Wigner factor is controlled mainly by the
difference $M_W-M_{W^*} > \Gamma_{W,{\rm tot}}$.
Because of this, the $M_{W^*}$ mass distribution
becomes quite wide, and the peak is shifted below the edge of phase
space by an amount  of order $2M_W-M_H$. 
Comparing the signal and background jet-jet invariant mass distributions, Figs.~\ref{fig:Hmjj} and \ref{fig:ggWWmjj}, we see that while the S/B ratio could be improved slightly by imposing a minimum $M_{jj}$, the loss of signal events would not lead to any overall improvement in the statistical significance of the signal. Note that we have not investigated further optimisation procedures, such as the cuts on the final-state lepton angles and azimuthal correlations between the quark jets, which may further improve the signal-to-background ratio.
%%%

It is obvious from Figs.~\ref{fig:ggWWmjj} and \ref{fig:Hmjj} that above the $WW$ threshold the situation becomes more favourable, since
the background contribution can now be reduced by requiring 
the invariant mass of the di-quark system to be close to $M_W$. Moreover, at higher Higgs masses
the mass resolution of the proton taggers is expected to improve \cite{RO}.

Returning to the case when $M_H < 2 M_W$, we have so far concentrated on the case when it is the {\it off-shell}
$W^*$ that decays hadronically, the on-shell $W$ decaying leptonically. There will of course be an equal number of signal events when this situation is reversed and the $W^*$ decays leptonically.\footnote{The ratio of fully hadronic, mixed hadronic-leptonic and fully leptonic decay channels for $H \to WW^*$ is rough $4:4:1$, with the mixed channel split evenly between the case when the $W$ or the $W^*$ decays leptonically.}  Therefore apart from around the threshold region 
$\sqrt{s} \sim 2 M_W$, the $M_{jj}$ distribution for the full
$q \bar  q l \nu$ sample of the signal will have a double-peak structure, corresponding to the superposition of a Breit-Wigner distribution peaked around $M_{jj} \sim M_W$ with a broader distribution peaked at lower mass, see Fig.~\ref{fig:Hmjj}.
The QCD $W$-strahlung process calculated above is only a background to the former component of the 
signal.\footnote{Actually, there will still be a QCD background contribution from $W^* q \bar q$ production with $W^* \to l \nu$ and $M_{q \bar q} \sim M_W$, but this will be very small.}
The only other potentially significant background contributions for the $M_{jj} \sim M_W$ case come from the
photon-photon fusion $\gamma\gamma\to WW^*$ subprocess and from the gluon-gluon fusion $ gg^{PP}\to WW^*$ subprocess
mediated by a quark loop.
As  discussed in Ref.~\cite{ww}, the former ($\gamma\gamma$) contribution can be strongly
(about 10 times) reduced
by imposing cuts on the forward proton  transverse
momenta, $p_t >100-200$~MeV/c.
Using the results of \cite{PM}, we estimate that the QCD quark box-diagram $ gg^{PP}\to WW^*$ contribution
is very small, on the level of  $1\%$ of the signal yield.
%
%On the other hand,
%by selecting such events we lose only about $40\%$ of the signal
%(purely leptonic or mixed lepton-quark) cross section.
Therefore, the statistical significance for these type of events
is very high:
for events with $p_t>100$~MeV/c the 
expected signal-to-background
ratio $S/B$ is of the order of 10.  This makes the channel
with the leptonic $W^*$-decay especially attractive when $M_H < 2 M_W$.
We expect that a detailed Monte Carlo simulation of both signal and
background events of this type would lead to a set
of signal-enhancing experimental cuts, reflecting the specific kinematics of these processes. 

In summary, we have considered in detail the exclusive production and decay to $WW$ or $W^*$ of a Higgs boson in conjunction with two forward protons at the LHC. We have focused on the $q \bar q l \nu$ final state, which constitutes just less than half the signal. For $M_H < 2 M_W$, there are two distinct scenarios in which either the $W$ or the $W^*$ decays leptonically. For the former, we identified and calculated the QCD background, showing it to be of similar magnitude to the (Standard Model) Higgs boson signal. We expect that the situation could be improved 
somewhat by optimising the cuts on the final-state particles. In the second scenario in which the $W^*$ decays leptonically, the background is expected to be very small. Overall, then this process offers a promising way of detecting the Higgs boson, either in the Standard Model or in particular supersymmetric extensions.

\section*{Acknowledgements}

We thank Mike Albrow, Brian Cox, Albert De Roeck, John Ellis, Jeff Forshaw,
Alan Martin, Sasha Nikitenko, Risto Orava, Krzystof Piotrzkowski and Georg Weiglein
for useful discussions. 
 MGR thanks the IPPP at the University of
Durham for hospitality. This work was supported by
the UK Particle Physics and Astronomy Research Council, by a Royal Society special
project grant with the FSU, by grant RFBR 04-02-16073
and by the Federal Program of the Russian Ministry of Industry, Science and Technology
SS-1124.2003.2.

\end{document}